# Characterizing HCI Research in China: Streams, Methodologies and Future Directions

**Tao Bi**
UCL Interaction Centre
University College London, London, UK
t.bi@ucl.ac.uk

**Yiyi Zhang**
Department of Social and Behavioural Sciences
City University of Hongkong, Hongkong
elly.zhang@my.cityu.edu.hk

**Chongyang Wang**
UCL Interaction Centre
University College London, London, UK
chongyang.wang.17@ucl.ac.uk

**Amid Ayobi**
UCL Interaction Centre
University College London, London, UK
amid.ayobi.14@ucl.ac.uk

## ABSTRACT

This position paper takes the first step to attempt to present the initial characterization of HCI research in China. We discuss the current streams and methodologies of Chinese HCI research based on two well-known HCI theories: Micro/Marco-HCI and the Three Paradigms of HCI. We evaluate the discussion with a survey of Chinese publications at CHI 2019, which shows HCI research in China has less attention to Macro-HCI topics and the third paradigms of HCI (Phenomenologically situated Interaction). We then propose future HCI research directions such as paying more attention to Macro-HCI topics and third paradigm of HCI, combining research methodologies from multiple HCI paradigms, including emergent users who have less access to technology, and addressing the cultural dimensions in order to provide better technical solutions and support.

1. INTRODUCTION

Human-computer Interaction (HCI) is an interdisciplinary research field involving multiple disciplines, such as computer science, psychology, social science and design. It studies the interaction between users and computer in order to better design technologies and solve real-life problems. HCI field has been increasingly popular, since the first conference on Human Factors in Computing Systems in 1982 in the United States[1]. With the remarkable growth of the Chinese technology market, HCI research receives increasing attention both in academia and industry, since the First International Symposium of Chinese Computer-Human Interaction (Chinese CHI) organized in 2013 [2].

In this position paper, targeting the workshop theme – HCI Research Highlights in China, we firstly discuss the mainstream HCI research topics that have been mainly investigated in China, basing on the theory of Micro/Macro-HCI [3]. Then we analyse the practice of applying HCI research methods in the Chinese HCI community, basing on the Three Paradigms of HCI [4]. The informal survey of Chinese publications at CHI 2019 [5] will be presented as an initial attempt to evaluate our positions. We contribute to the workshop discussion by suggesting several potential directions for HCI research in China. For instance, there should be more attention to Macro-HCI topics and third paradigm of HCI. Combining research methodologies from multiple HCI paradigms are encouraged to achieve broader and deeper understanding and research validity. We also propose that Chinese HCI researchers should focus more on the emergent user groups that have limited accessibility to technology. We specifically state that HCI research in China should pay more attention to the study of cultural factors when designing new technologies, taking the example of the digital red-packet [6]. These will be further explained in the following sections.

2. STREAMS OF HCI RESEARCH

The research conducted in the Chinese HCI community is slightly different compared with the international HCI research tradition. In China, HCI research in both academia and industry study tends to be more technology-focused with emphasis on back-end hardware building or software engineering, such as speech recognition interfaces and Chinese keyboard applications for mobile phones. Researchers develop new or improve existing interfaces by quantifying how much faster or more accurately users can complete tasks. This type of HCI research is relatively technology-driven and focused on the computer-orientated quality attribute s such as performance, robustness, reliability, efficiency, scalability, etc.

With the increasing popularity of user-centred design (UCD) approach, more industry companies (e.g., Alibaba, Tencent) have raised their awareness on the usability of their products and services in order to provide better user experiences. However, there is still a lack of attention to the human behaviour surrounding computer systems. Much less attention has been paid to the behavioural and cultural dimensions of human-computer interactions. Protentional questions could be explored including the troubles people face in everyday life when using technologies and their individual goals and preferences. A better understanding of individual practices is key to shape more usable and useful technologies [7].

Shneiderman [3] proposes the two categories of HCI research: Micro-HCI and Macro-HCI. Chinese HCI research seems to focus more on the Micro-HCI category that aims to improve the completion time and performance of tasks. The current international HCI community investigates also the Macro-HCI category that involves broader research topics, such as motivation, social collaboration, trust and wellbeing, engagement, as well as the emotional experience. These Macro-HCI research questions have more societal-level impact and they need to be studied with research methods borrowed from sociology and psychology fields (i.e. qualitative research rather than controlled lab experiments). Chinese HCI research seems to focus more on building new artefacts and empirically evaluating the performance of HCI system in a quantitative way. Based on these two categories, we conducted an informal survey of the 22 full papers from Chinese authors (in total, 25 papers published at CHI 2019) (See **Figure 1.**). Only two papers are addressing Macro-HCI topics such as the behaviour and needs of blind people doing online shopping and using mobile technology for daily health issues. The rest 20 papers are about Micro-HCI topics such as improving screen-reader keyboard, full hand interaction on Smartphone, and virtual reality head-mounted Displays. We encourage to examine not only Micro-HCI but also Macro-HCI research topics in order to solve practical human issues.

Artificial intelligence (AI) research also becomes increasingly popular. Current AI researches (i.e. Deep Learning, Computer Vision, Natural Language Processing, etc.) in China lay a particular emphasis on developing the intelligence of the machine with the research focuses on pushing the boundaries of the model performance. Indeed, dealing with big data from the large-scale population is easier than deeply understanding the individual. The initial purpose of AI is to better serve human being. In china, there is an undermined research area of understanding the human side, including research questions such as where AI is needed, what questions should be addressed by AI techniques, and more importantly, how AI should be designed in the life of human individuals, and how AI should be user-centred. HCI research, therefore, should play an important role in the development of AI especially in China, to solve real-necessary human and societal problems.

## 3. METHODOLOGIES

Steve Harrison etc. [4]. introduced the concepts of 'three paradigms of HCI' as a well-known approach for characterizing HCI research (See Table 1, 2, 3 as an overview of The three paradigms of HCI). The first-paradigm HCI research focuses on optimizing man-machine fit (e.g. reduce human errors), and design works as problem solving is the main goal in the first paradigm. The HCI research in the second paradigm optimizes the communication and information transfer between machine and human (e.g. by considering human's cognitive load). The appropriate research questions are how to model users and design to facilitate the communication. It is often constructed as hypothesis-testing that happens in controlled conditions with the methods (statistical analysis, classification, corroboration) from computer science and laboratory behavioural sciences in order to understand the interaction. In second paradigm, researchers often value a clean or well-defined principle as the generalization of the study findings. Therefore, if there is a defined problem or a constrained project, the approaches from the first and second paradigm will be useful to guide the design solutions.

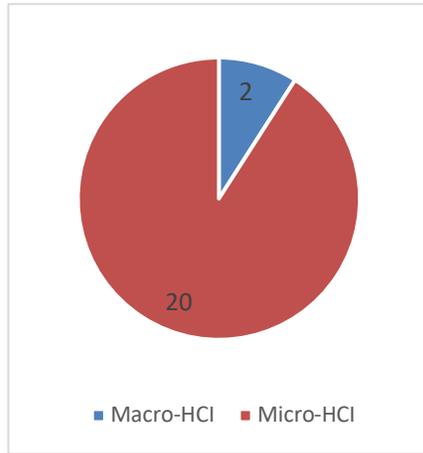

**Figure 1. HCI Research Streams in China: Survey of Chinese publications at CHI 2019**

**Table 1: HCI Paradigm 1 - Human Factors [4]**

| Metaphor | Interaction as human-machine coupling |
|---|---|
| Goal | Solving problems during interaction |
| Questions | How to fix problems? |
| Disciplines | Engineering, programming, ergonomics |
| Values | Reduce errors even with Ad hoc |

**Table 2: HCI Paradigm 2 – Classical Cognitivism/Information Processing [4]**

| Metaphor | Interaction as Information communication |
|---|---|
| Goal | Optimizing accuracy and efficiency |
| Questions | - What are mismatches in communication between computers and people?<br>- How to accurately model what people do?<br>- How to improve the efficiency? |
| Disciplines | Laboratory and behavioural sciences |
| Values | Optimization; Generalizability; Principled evaluation; Structured design; Reduction of ambiguity; Top-down view of knowledge |

Table 3: HCI Paradigm 3 - phenomenologically Situated [4]

| Metaphor | Phenomenologically situated interaction |
|---|---|
| Goal | Supporting situated interaction in real world |
| Questions | - What existing situated activities in the world should we support?<br>- How do users appropriate technologies?<br>- How can we support interaction without constraining it too strongly by what a computer can do or understand?<br>- What are the politics and values at the site of interaction, and how can we support those in design? |
| Disciplines | Ethnography, ethnomethodology, action research, interaction analysis |
| Values | Goal is to grapple with the full complexity around the system. What goes on around systems is more interesting than what's happening at the interface. What you don't build is as important as what you do build. Construction of meaning in interaction activity is important. |

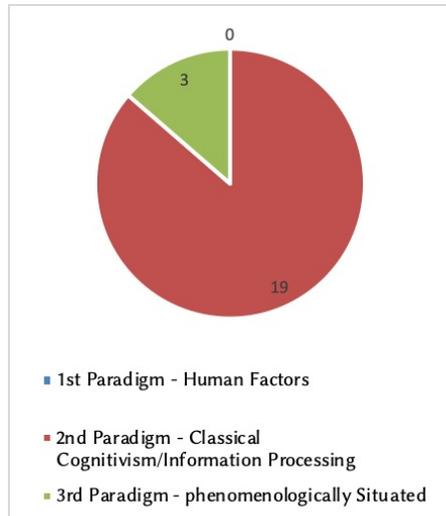

Figure 2: An informal survey of 22 full papers from China published at CHI 2019

Although the generalized principle from the data are useful for giving insights of understanding the interaction, it is a bit regardless of the complicated real-life context and not sufficient to directly explain the real-world situated behaviours and activities that are always complicated and messy. In order to better understand what users are doing and what people really need we need specifically analyse the situated behaviours during the interaction. The third paradigm, phenomenologically situated paradigm, centrally focuses on meaning and meaning creation based on the human experience and from multiple perspectives. Using the methods from field studies, Ethnography, ethnomethodology, action research, interaction analysis provides a more detailed and richer descriptions tied to specific situations [8].

In Chinese HCI research community, 2nd paradigm have be dominant in HCI research, while the third paradigm, phenomenologically-situated, deserves more attention. An informal survey (See **Figure 2**) of the 22 full papers from Chinese authors (in total, 25 papers published at CHI 2019) shows only 3 papers are thought as situated perspective paradigm (See Table 1.) while 18 papers are classical cognitivism/information processing paradigm and 1 paper is from human-factor paradigm.

We are not arguing that third-paradigm HCI research is right, while first and second-paradigm HCI research are wrong. Indeed, they are equally important. Rather, we are arguing that the third-paradigm HCI research deserves more attention in the Chinese research community. Therefore, we suggest that it would be valuable to conduct more HCI research or using methods from the third paradigm or of combination of multiple paradigms. This will induce not only different types of questions but also the different methods for answering them, which is beneficial for a deeper and broader understanding of interaction and the phenomenon especially for research involved with social and cultural factors. For instance, "why people like using WeChat than QQ?" These types of research are always highly involved with social and cultural factors that plays a significant role in Chinese society. In addition, approaches from multiple paradigm provide a stronger validity of the research, which can more efficiently identify the solid contribution to Chinese HCI research community. By introducing third paradigm HCI research methods may cause conflicts between researchers because methods from one paradigm may sounds not valid to some researchers from another paradigm. This may be because the difference of their understanding of how it is generated. However, HCI itself is an interdisciplinary subject so that they should learn from each other and to maintain the academic solidness and rigor.

## 4. TOWARDS INCLUDING EMERGENT USERS

In spite of the rapid growth of economy during the past decade, China still remains a developing country that faces various challenges and a series of societal issues such as the inequality in Wealth and Education Resources Distribution, as well as the increasingly serious aging issue. International

HCI community has started to include people experiencing these issues. The HCI notion – "emergent users" refers to the people who received less education or have a low level of living conditions [9]. This also includes people who are living in the underdeveloped rural areas laggard in infrastructure, education and economic construction [10].

Emergent users are different compared with the traditional users in terms of their needs and experiences. There are 85 million disabled people in China [11]. Chinese HCI researchers started to focus on this group, which can be identified from Chinese CHI 2019 publication (two full papers targeting the design for visually impaired people). However, not all of them has the access to the advanced assistive technology. Some disabled people who live in a poor financial condition cannot afford advanced assistive technology (e.g. a smart wheelchair). Similarly, in some undeveloped regions of China, the children who do not have the supervisions from parents could easily get addicted to unhealthy online information and mobile games through the widespread smart phones. The question will be who should be responsible for the addition of kids to mobile games and the impact on their development? There have been many systems designed for "Anti-addition system". However, why are they not working well. HCI researchers, therefore, should really understand the real problem of these group of people from a perspective of social, cultural, ethical, or economical contexts. Research should look at how technology could be introduced to emergent users at young age when they are just beginning to appropriate advanced mobile devices, in order to avoid the potential negative influences, such as addictions.

Therefore, the development of Chinese HCI industry should not only focus on the novelty of trending technology such as fancy and more advanced interfaces, but more importantly take into account the realistic utility of the emerging technology for better practical support to the demanding needs derived from particular social problems. HCI researchers should also look at how to design assistive applications and devices that are affordable, practical and sufficient for emergent users. To achieve these, it is essential to understand the living and economic conditions of respective groups from a sociological perspective. Research methodologies in sociology could have potential for digging into deeper social problems, in order to provide relevant and feasible technical solutions to specific issues. In addition, these types of HCI research can also help the policymakers to inform better regulations, public policy and even laws for these population, especially in the areas about accessibility, ergonomics, health, and education [12].

5. TOWARDS ACKNOWLEDGING CULTURE DIMENSIONS.

As HCI research is heavily involved with and affected by cultural factors, Chinese HCI community should study more on the cultural and behavioural factors to better understand the people's needs and behavioural differences when designing new HCI system. China is a country where culture plays an important role in shaping the human behaviours and citizen's lifestyle. Importantly, Chinese culture is very diverse. People from different region or ethnic groups may have different preferences and requirements for the human-computer interaction. Previous Chinese HCI conferences organizer also identifies the importance of looking at the difference between ethnic groups in China when considering the user behaviours [13].

To better understand the cultural factor, here, we rethink the example of digital application of Red-Packet. Red-Packet (also called Hongbao in Madarin) is a nationwide tradition during the Chinese New Year festival [14]. Older people normally prepare a red envelope containing lucky money and give it to young children as a gift with good wishes. In 2015, WeChat (an instant message app developed by the Chinese technology company Tencent) firstly introduced the digital version of Red-Packet that allows people to give and receive the lucky money via online message. WeChat also allows people to "grab red-packets" within their group chat. During the Chinese New Year festival, people also have the chance for winning 3rd-party red-packets by "shaking their phones ceaselessly"[6]. This new format of Red-Packet, as a novel way to do social interaction, became very popular and caused a huge amount of participation (around 10 billion) especially during the New Year Eve.

As HCI researchers, we need to think about people's motivation for giving/receiving Red-Packet, while other questions need to be considered include how the digital red packet changes/exists with the traditional red-packet culture? and how digital red-packet is differing from traditional red-packet? Why does the traditional one is still happening? Why do people want to grab/fight for Red-Packet within their group chat? An in-depth understanding of these cultural behaviours will help to maintain and even promote the engagement of this cultural tradition, as well as enhance the social relationships between people.

## 6. CONCLUSION

In this position paper, we have outlined the current landscape of research streams and applied methodologies in the Chinese HCI community. Based on this understanding, we propose that the Chinese HCI community should not only focus on incrementally innovating hardware to improve the quality attributes of HCI systems (Micro-HCI), but also investigate the behavioural and cultural dimensions of human-computer interactions to explore fundamental question, such as: what are the needs of the explored user groups and how could these be supported with technologies (Macro-HCI)?In order to better address these questions, it seems reasonable to put more emphasis on, firstly, understanding a phenomenon before developing the interface, and to adopt more research methods from the third paradigm of HCI (phenomenologically situated interaction). Combination of multiple HCI paradigms are encouraged. More importantly, the Chinese HCI community should pay attention to designing for emergent users who have less access to advanced technologies. Last but not least, cultural factors are particularly important and should be carefully looked at when conducting HCI research in China.